\documentclass[
    ,final            
  ]
  {aipproc}

\layoutstyle{6x9}

\usepackage{graphicx,multirow}
\usepackage{axodraw}
\usepackage{pstricks}
\usepackage{color}
\makeatletter
\newcounter{subsubsubsection}[subsubsection]
\def\subsubsubsectionmark#1{}

\def\subsubsubsection{\@startsection
     {subsubsubsection}{4}{\z@} {-3.25ex plus -1
     ex minus -.2ex}{1.5ex plus .2ex}{\normalsize\it}}
\def\l@subsubsubsection{\@dottedtocline{4}{4.8em}
     {4.2em}}
\newcounter{subsubsubsubsection}[subsubsubsection]
\def\subsubsubsubsectionmark#1{}

\def\subsubsubsubsection{\@startsection
     {subsubsubsubsection}{5} {\z@} {-3.25ex plus -1
     ex minus -.2ex}{1.5ex plus .2ex}{\normalsize\bf}}
\def\l@subsubsubsubsection{\@dottedtocline{5}
     {5.8em}{5.2em}}

\def\>{\rangle}
\def\<{\langle}

\def\P{{\cal P}}

\def\Lb{\Lambda_b}

\def\Ls{\Lambda}
\def\={&=&}
\def\<{\langle}
\def\>{\rangle}

\def\beq{\begin{equation}}
\def\eeq{\end{equation}}

\def\beqy{\begin{eqnarray}}
\def\eeqy{\end{eqnarray}}
\def\beqynn{\begin{eqnarray*}}
\def\eeqynn{\end{eqnarray*}}

\begin{document}

\title{\bf Lepton Asymmetries in FCNC Decays of $\Lb$}

\author{L. Mott}{
address={Department of Physics, Florida State University, Tallahassee, FL 32306}
}

\author{W. Roberts}{
address={Department of Physics, Florida State University, Tallahassee, FL 32306}
}

\begin{abstract}
Lepton polarization asymmetries for the flavor-changing neutral current (FCNC) dileptonic decays of $\Lb$ baryons are calculated using single-component 
analytic (SCA) and multi-component numerical (MCN) form factors. We show that these polarization asymmetries are insensitive to the transition form factors 
and, thus, the effects of QCD in the nonperturbative regime. Therefore, these observables can provide somewhat model independent ways of extracting the 
Wilson coefficients.
\end{abstract}


\maketitle

\section{Introduction}

When investigating the flavor-changing neutral current (FCNC) decays of heavy hadrons, several experimentally measurable quantities contain valuable
information about the Wilson coefficients that enter into the effective Hamiltonian that describes these decays. Among these are observables involving
the final state leptons. These lepton asymmetries are important because, being defined as ratios of decay rates, they are expected to be less
sensitive to the nonperturbative QCD dynamics. Thus, they should offer somewhat model independent ways of determining the values of the Wilson
coefficients.

In this work, we present numerical results for longitudinal lepton polarization asymmetries (LLPAs) for the baryonic FCNC process
$\Lb\to\Ls^{(*)}\ell^+\ell^-\,\,\,\,(\ell=\mu,\tau)$. These asymmetries are calculated using the SCA and MCN form factors computed in
\cite{morob}.

\section{Results}

\begin{table}
\caption{Integrated longitudinal lepton polarization asymmetry for $\Lambda_b\rightarrow \Lambda^{(*)}\mu^{+}\mu^{-}$ in units of $10^{-2}$. The
numbers in the column labeled SM1 are obtained using the SCA form factors with standard model Wilson Coefficients. The numbers in the column labeled
SM2 are also obtained using SM Wilson coefficients, but with the MCN form factors. The numbers in the column labeled SUSY are obtained using the MCN
form factors with Wilson coefficients from a supersymmetric scenario. The column labeled LD refers to the long distance contributions of the
charmonium resonances, with `a' indicating that these contributions have been neglected, and `b' indicating that they have been included. In this
table, it is assumed that the $\Lambda(1600)$ is the first radial excitation.}
{\begin{tabular}{cccccc}
\hline\hspace{12pt} State, $J^{P}$\hspace{12pt}       &\hspace{12pt} LD\hspace{12pt} &\hspace{12pt} SM1\hspace{12pt} &\hspace{12pt} SM2\hspace{12pt}
&\hspace{12pt} SUSY\hspace{12pt} &\hspace{12pt} Chen {\it et al.} \cite{chen2}\hspace{12pt} \\
& & & & &\small QCDSR $\,\,$\small PM  \\ \hline
$\Lambda(1115)\,1/2^{+}$    & a & $-58.1$ & $-58.5$ & $-53.2$ & $-58.3\,\,\,\,\,\,\,\,\,\,\,-58.3$ \\
                            & b & $-51.6$ & $-51.9$ & $-49.1$ &$-$ \\ \hline
$\Lambda(1600)\,1/2^{+}$    & a & $-45.4$ & $-45.6$ & $-41.4$ & $-$ \\
                            & b & $-40.1$ & $-40.4$ & $-37.1$ & $-$ \\ \hline
$\Lambda(1405)\,1/2^{-}$    & a & $-49.7$ & $-50.2$ & $-45.6$ & $-$ \\
                            & b & $-43.9$ & $-44.4$ & $-41.4$ & $-$ \\ \hline
$\Lambda(1520)\,3/2^{-}$    & a & $-46.7$ &  $-47.5$ & $-43.1$ & $-$ \\
                            & b & $-41.4$ & $-42.1$ & $-38.8$ & $-$ \\ \hline
$\Lambda(1890)\,3/2^{+}$    & a & $-38.1$ & $-38.3$ & $-34.8$ & $-$  \\
                            & b & $-34.1$ & $-34.2$ & $-30.3$ & $-$ \\ \hline
$\Lambda(1820)\,5/2^{+}$    & a & $-38.7$ & $-39.6$ & $-36.0$ & $-$  \\
                            & b & $-34.3$ & $-35.0$ & $-31.7$ & $-$ \\ \hline
\end{tabular}\label{llpa1}}
\end{table}

\begin{table}
\caption{Integrated longitudinal lepton polarization asymmetry for $\Lambda_b\rightarrow \Lambda^{(*)}\tau^{+}\tau^{-}$ in units of $10^{-2}$. The
columns 
are labeled as in Table \ref{llpa1}.}
{\begin{tabular}{ccccccc}
\hline\hspace{12pt} State, $J^{P}$\hspace{12pt}      &\hspace{12pt} LD\hspace{12pt} &\hspace{12pt} SM1\hspace{12pt} &\hspace{12pt} SM2\hspace{12pt}
&\hspace{12pt} SUSY\hspace{12pt} &\hspace{12pt} Chen {\it et al.} \cite{chen2}\hspace{12pt} \\
& & & & &\small QCDSR $\,\,$\small PM    \\ \hline
$\Lambda(1115)\,1/2^{+}$   & a & $-10.7$ & $-10.8$ & $-8.2$ & $-10.9\,\,\,\,\,\,\,\,\,\,\,-10.9$ \\
                           & b & $-10.2$ & $-10.3$ & $-8.1$ & $-$  \\ \hline
$\Lambda(1600)\,1/2^{+}$   & a & $-3.4$ & $-3.5$ & $-2.6$ & $-$ \\
                           & b & $-3.1$ & $-3.1$ & $-2.5$ & $-$  \\ \hline
$\Lambda(1405)\,1/2^{-}$   & a & $-5.9$ & $-6.0$ & $-4.5$ & $-$  \\
                           & b & $-5.5$ & $-5.6$ & $-4.4$ & $-$  \\ \hline
$\Lambda(1520)\,3/2^{-}$   & a & $-4.6$ &  $-4.6$ & $-3.4$ & $-$  \\
                           & b & $-4.2$ & $-4.2$ & $-3.2$ & $-$  \\ \hline
$\Lambda(1890)\,3/2^{+}$   & a & $-0.81$ & $-0.80$ & $-0.58$ & $-$  \\
                           & b & $-0.54$ & $-0.52$ & $-0.46$ & $-$  \\ \hline
$\Lambda(1820)\,5/2^{+}$   & a & $-1.4$ &  $-1.4$ & $-0.99$ & $-$  \\
                           & b & $-1.1$ & $-1.1$ & $-0.89$ & $-$  \\ \hline
\end{tabular}\label{llpa2}}
\end{table}

\begin{figure}
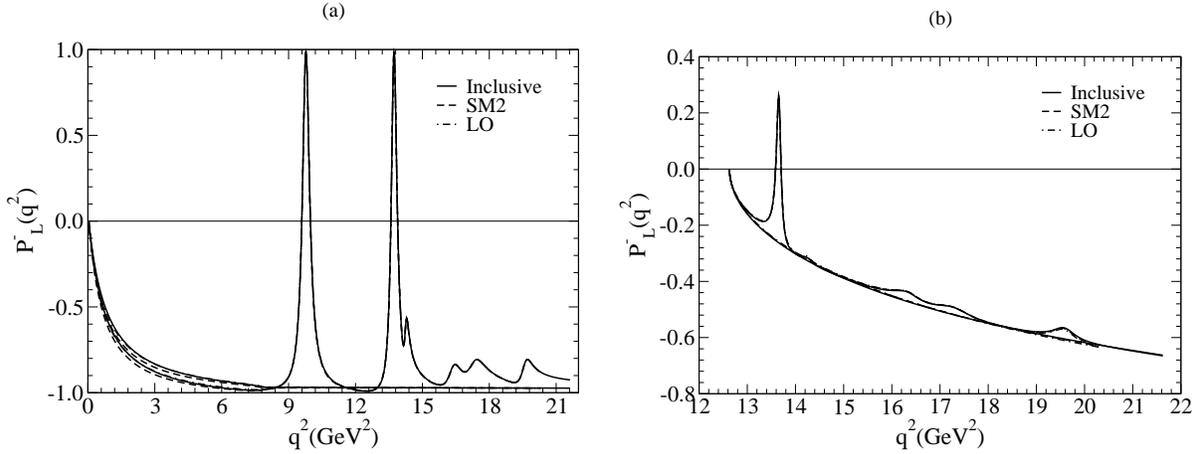

\centerline{\includegraphics[width=3.0in]{111_Mott-f1.eps}\,\,\,\,\,\,
\includegraphics[width=3.0in]{111_Mott-f2.eps}}
\caption{$\P_L^-(q^2)$ for (a) $\Lambda_{b}\rightarrow\Lambda(1115)\mu^{+}\mu^{-}$ and (b) $\Lambda_{b}\rightarrow\Lambda(1115)\tau^{+}\tau^{-}$
without 
and with long distance (LD) contributions from charmonium resonances. The solid curves arise from the free quark process $b\to s\ell^+\ell^-$, while
the 
dashed curves are the SM2 results, and the dot-dashed curves are the leading order SM results..}\label{fig:PLq212p}
\end{figure}

The integrated LLPAs we obtain are shown in Tables \ref{llpa1} and \ref{llpa2}. Each table displays the results for the SM calculations using two
models for the form factors, as well as one SUSY scenario with the MCN form factors. In addition, results obtained omitting and including the long
distance (LD) contributions are presented. We also compare our results with those of Chen {\it et al.} \cite{chen2}, where they have used QCD sum
rules (QCDSR) and a pole model (PM) for the form factors. As can be seen, the results from both SM1 and SM2 are in agreement with each other and the
QCDSR and PM results reported in \cite{chen2} for both channels. For decays to the excited states, the SM1 and SM2 results are in agreement for both
channels as well. In the SUSY scenario we use, where $C_7^{SUSY}=-C_7^{SM}$, the LLPAs are quite different from the SM results.

In Fig. \ref{fig:PLq212p}, we show the differential LLPA for decays to the ground state for the SM2 case. In addition, we also display results for
the free-quark process $b\to s\ell^+\ell^-$ as well as the leading order results with no form factor dependence. It can be seen that these curves are
nearly indistinguishable. This is clear evidence that the LLPAs can be considered as model independent quantities.

\section{Outlook}

We have shown that the LLPAs are largely independent of the form factor model chosen. We have also shown that they can be sensitive to new physics.
Thus, LLPAs can be useful in determining the values of the Wilson coefficients and in looking for new physics beyond the SM. In addition to the
longitudinal component, there are also transverse and normal components of polarization. These may also prove to be less sensitive to the form
factors, yet sensitive to new physics. In addition to the lepton asymmetries, there are also baryon polarization asymmetries that can be studied.
Since baryonic decays could maintain the helicity structure of the effective Hamiltonian, baryon asymmetries may also be sensitive to beyond
the SM scenarios.

\section*{Acknowledgment} We gratefully acknowledge the support of the Department of Physics, the College of Arts and Sciences, and the Office of 
Research at Florida State University. This research is supported by the U.S. Department of Energy under contract DE-SC0002615.

\bibliographystyle{aipproc}

\end{document}